\documentclass[conference]{IEEEtran}
\IEEEoverridecommandlockouts

\usepackage{cite}
\usepackage{amsmath,amssymb,amsfonts}
\usepackage{acronym}
\usepackage{graphicx}
\usepackage{textcomp}
\usepackage{xcolor}
\usepackage{threeparttable}
\usepackage{booktabs}
\usepackage[hidelinks]{hyperref}
\usepackage{multirow}

\usepackage{pifont}
\usepackage{subcaption}

\def\BibTeX{{\rm B\kern-.05em{\sc i\kern-.025em b}\kern-.08em
T\kern-.1667em\lower.7ex\hbox{E}\kern-.125emX}}

\newcommand{\cmark}{\ding{51}}%
\newcommand{\xmark}{\ding{55}}%
\newcommand{\jj}{\jmath}

\acrodef{cig}[CIG]{converter-interfaced generator}
\acrodef{pll}[PLL]{phase-locked loop}
\acrodef{sm}[SM]{synchronous machine}
\acrodef{vsm}[VSM]{virtual synchronous machine}

\begin{document}

\title{On the Emulation of Synchronous Machine Dynamics by Converter-Interfaced Generators
\thanks{G. Tzounas is supported by the Swiss National Science Foundation under NCCR Automation (grant no.~51NF40 18054). 
F. Milano is supported by the Sustainable Energy Authority of Ireland~(SEAI), under the project FRESLIPS (grant no.~RDD/00681).}
}

\author{\IEEEauthorblockN{Georgios Tzounas, {\em IEEE Member}}
\IEEEauthorblockA{\textit{Power Systems Laboratory} \\
\textit{ETH Z{\"u}rich}\\
Z{\"u}rich, Switzerland \\
georgios.tzounas@eeh.ee.ethz.ch}
\and
\IEEEauthorblockN{Federico Milano, {\em IEEE Fellow}}
\IEEEauthorblockA{\textit{School of Electrical \& Electronic Engineering} \\
\textit{University College Dublin}\\
Dublin, Ireland \\
federico.milano@ucd.ie}
}

\maketitle

\begin{abstract}
This paper discusses the conditions that a device needs to satisfy  to replicate the behavior of a conventional \ac{sm} connected to a power network.  The conditions pertain to the device's stored energy, time scale of response,  oscillation damping, and behavior during short-circuits.  Relevant remarks for devices that do/don't satisfy these conditions are discussed through an illustrative numerical example as well as through simulation results based on a modified version of the well-known WSCC 9-bus test system.
\end{abstract}


\section{Introduction}

\subsection{Motivation}

Unlike \acfp{sm}, \acp{cig} do not inherently provide inertia to the power grid, are often stochastic, and operate with small or no power reserves \cite{mauricio2009frequency}.  These properties pose serious challenges to the transition from a \ac{sm}- to a \ac{cig}-dominated power system \cite{milano2018foundations}.  On the other hand, the behavior of \acp{cig} is dictated by their control loops, and hence these resources are very flexible, since they can be designed using a broad range of control strategies to provide fast and effective regulation \cite{unruh2020overview, zhong2022improving}. 

\subsection{Literature Review}

The key role that \acp{sm} play in the dynamic performance of power systems is highly appreciated by system operators.  This has motivated important efforts for the design of \ac{cig} control methods able to offer the auxiliary services conventionally provided by \acp{sm}, including inertial response, voltage and frequency regulation, and suppression of electromechanical oscillations.  The application of these methods varies from the control of a single power electronic converter to the controlled aggregation of multiple heterogeneous converter-based resources \cite{pudjianto2007virtual, edgeflex:d21, 2021:tzounas}.  Moreover, a part of these methods has explicitly aimed to replicate the dynamic response of \acp{sm}, which has led to the concept of \acf{vsm}.  The development of \ac{vsm} is still in an early stage and various implementations have been proposed in the recent literature,
for example, we cite \cite{driesen2008virtual, d2015virtual, cheema2020comprehensive}.

In a different vein, several recent studies have proposed analogies of \acp{sm} with different kinds of devices, and with a goal to study various problems, including frequency control, synchronization of power converters, transient stability, etc.  For example, the authors in \cite{d2013equivalence,liu2015comparison} propose that a droop control is equivalent to a \ac{vsm}, whereas in \cite{8959148}, the authors suggest an equivalence between a \ac{sm} and a grid-forming converter.  In \cite{taul2019overview}, it is suggested that a \ac{pll} used for converter synchronization is analogous to a \ac{sm}.  Yet another analogy is outlined in \cite{dorfler2012synchronization}, where a non-uniform Kuramoto oscillator is described as equivalent to an overdamped \ac{sm}.

Motivated by the above line of works, this paper discusses the validity of characterizing a non-synchronous device as equivalent to a \ac{sm}. Such characterization, apart from a formal mathematical equivalence, should depend also upon a set of additional and critical constraints. A qualitative summary of these constraints as well as of the implications of their violation is complementary to the existing literature and can provide a didactic value for researchers working on the design of \ac{cig} control methods.

\subsection{Contribution}

The contributions of the paper are as follows:

\begin{itemize}
\item A qualitative description of the conditions that make a power electronic-based device behave like a traditional \ac{sm} connected to a power system.  These conditions pertain to the device's energy availability, time scale of action, damping, and response to short-circuits.
\item A discussion on the ability to satisfy these conditions of devices proposed in the literature as equivalent \acp{sm}, including \acp{vsm}, droop controllers and \acp{pll}.
\end{itemize}

\subsection{Organization}

The remainder of the paper is organized as follows.  Section~\ref{sec:secondorder} recalls the mathematical analogy between a generic second-order oscillator and the classical \ac{sm} model.  The requirements that a device needs to fulfill to replicate the behavior of a traditional \ac{sm} are presented in Section~\ref{sec:conditions}.  The case study is discussed in Section~\ref{sec:case}. Finally, conclusions are drawn in Section~\ref{sec:conclusion}.

\section{Synchronous Machines as Oscillators}
\label{sec:secondorder}

Let us recall the 
classical \ac{sm} model \cite{kundur:94}:
\begin{equation}
  \label{eq:motion}
  \begin{aligned}
    \dot\delta &= \Omega_b \, \omega \, , \\
    2H \, \dot\omega &= p_m - p_e(\delta) - D \, \omega \, , 
  \end{aligned}
\end{equation}
where $\delta$ (rad) is the rotor's angle and $\omega$ (pu) the rotor's speed variation with respect to the reference angular frequency; $\Omega_b$ (rad/s) is the synchronous frequency; $H$~(s) is the \ac{sm} inertia constant and $D$ its damping factor; $p_m$ and $p_e(\delta)$ are, respectively, the \ac{sm} mechanical and electrical power output in pu, with $p_e(\delta) = {e' v} \sin(\delta - \theta)/{X}$, where $e'$ is the \ac{sm} internal emf; $\bar v = v \angle \theta$ is the voltage at the \ac{sm} terminal bus; and $X$ is defined as the sum of the \ac{sm} transient reactance and the reactance that connects the \ac{sm} to its terminal bus.

\begin{figure}[ht!]
  \centering
  \resizebox{0.59\linewidth}{!}{\includegraphics{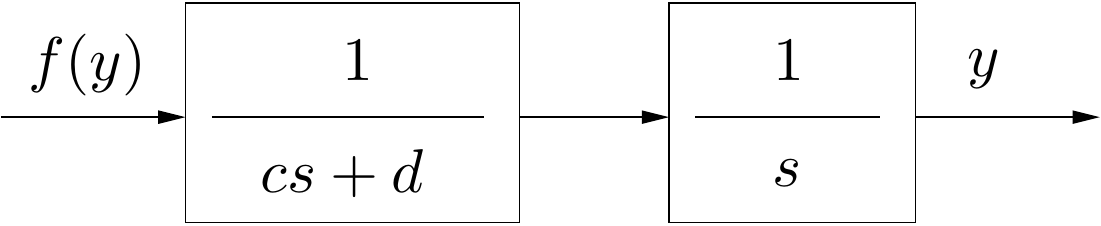}}
  \caption{Block diagram of \eqref{eq:second}.}
  \label{fig:block}
\end{figure}

Let us rewrite \eqref{eq:motion} as follows:
\begin{equation}
  \label{eq:second}
  c \, \ddot y + d \, \dot y - f(y) = 0 \, , 
\end{equation}
where $y\equiv\delta$, $c = 2H$, $d=D$,
$f(y)=\Omega_b(p_m-p_e(\delta))$.  The last equation describes a very well-known concept, i.e.~the \ac{sm} is a second-order oscillator, where the damping is determined by $D$ and the ``reluctance'' to allow frequency variations is quantified by $H$.
The block diagram of \eqref{eq:second} is depicted in Fig.~\ref{fig:block}, where $s\in \mathbb{C}$ is the complex Laplace frequency.  

The literature abounds of variants of \eqref{eq:second}, such as the Van der Pol \cite{7336530} and the Li{\'e}nard-type oscillator \cite{8263726}, with additional non-linear terms, e.g.~$d = g(y)$, and/or forced input oscillations.  We acknowledge but do not discuss these models as they can be considered to be part of the broader category of synchronization mechanisms, such as \acp{pll}.  Moreover, the very same model shown in \eqref{eq:second} is ubiquitous in a broad class of engineering systems, which (or even more often, parts of which) can be reasonably approximated by a suitable second-order system of the same shape.  Then, taking into consideration the importance of \acp{sm} in power systems, one may observe a striking equivalence of the \ac{sm} with, in principle, any other system that can be described in the same form, e.g.~with a given second-order automatic controller.  

The main concept discussed in this paper is that, contrary to what is underlying (to lower or higher extent) in some recent works, a given device expressed in the form of \eqref{eq:second} is, in general, not equivalent to a \ac{sm}.  The only obvious analogy between such a device and a \ac{sm} is that they are both \textit{special cases} of the same, broad family of oscillators.
Moreover, a given device can be considered as emulating the behavior of a traditional \ac{sm} connected to a power network, if and only if a set of additional constraints are met.  These constraints are duly discussed in the next section.

\section{Synchronous Machine Emulation}
\label{sec:conditions}

In this section we discuss the conditions that a device needs to satisfy so that it replicates the behavior of a conventional \ac{sm} connected to a power grid.  These conditions pertain to the availability of energy, the time scale of response, the behavior during short-circuits,
and the damping of oscillations.

\subsection{Time Scale}
\label{sec:timescale}

The time scale of the dynamic response of the emulating device must be similar to that of a \ac{sm}.  A typical range of the inertia time constant $H$ in a \ac{sm} is [2,10]~MWs/MVA. The value of $H$ has a physical meaning and represents the time (in seconds) for which the \ac{sm} could inject its rated power to the system if disconnected from its turbine.  Therefore, second-order oscillators in the form of \eqref{eq:second} that respond in a different time frame do not resemble the behavior of a \ac{sm}.

\subsection{Stored Energy}
\label{sec:energy}

In a \ac{sm} of rated power $S_n$~MVA, the rotating mass has in nominal conditions stored kinetic energy $H S_n$~MWs.  After a negative (positive) mismatch between the mechanical power $p_m$ and electrical power $p_e$ and until primary regulation is initiated, this physical storage is the crucial mechanism that maintains the system's power balance but also the \ac{sm} synchronism, by decreasing (increasing) instantaneously its stored energy as the rotor decelerates (accelerates).  Maintaining synchronism and power balance are inextricable in a \ac{sm}, and hence, a \ac{sm}-emulating device is also required to include mechanisms that account for both tasks.

Regarding the power balance, a device that emulates a \ac{sm} should have sufficient stored energy that is available very fast (ideally instantaneously) after a power mismatch $\Delta p = p_m-p_e \neq 0$ occurs.  ``Very fast'' in this context basically refers to the time delay between the occurrence of the disturbance and the initiation of the device's response.  For \acp{cig}, instantaneous (i.e.~delay-free) provision of the required energy during an imbalance is not possible, and so the condition may be relaxed to a requirement for a very fast response.  This however may raise concerns, as it leads to a time window right after the disturbance that remains uncovered \cite{milano2018foundations}.  Overall, energy storage is by no means a trivial requirement for a \ac{sm}-equivalent device.

\subsection{Oscillation Damping}
\label{sec:damping}

Not well-damped oscillations are undesired. Thus, in an emulation of a \ac{sm} where the damping is a fully controlled parameter, it is reasonable that one decides to remove oscillations during the design (e.g., in the case of \eqref{eq:second}, by choosing a large $d$). 
We recall that \acp{sm} are designed for high efficiency and thus include a relatively small damping.  A typical range of $D$ for \eqref{eq:motion} is [2,3]~pu to account both for mechanical damping and effect of damper windings. In higher-order (e.g.~sixth-order) machine models, the effect of damper windings is explicitly represented in the model and thus $D$ can be chosen lower or even zero.   Then, poorly damped electromechanical oscillations are partially suppressed by some form of damping control, but the resulting response is still oscillatory.  In theory, good damping of \ac{sm} oscillations could be achieved through prime movers, but the latter are not fast-enough due to mechanical constraints. Thus, in practice, oscillations are damped through the \ac{sm} excitation system, but the effect is limited due to the weak coupling  of voltage with power angle.

The above problems do not exist in power electronic-based devices, which can provide a fast response and thus also be designed for very good damping.  However, it is worth noting that, an overdamped response, even if desired, does not replicate the conventional behavior of a \ac{sm} connected to a power network.

\subsection{Link of Time Scale with Energy and Damping}

A qualitative way to study the link of time scale with energy and damping in model \eqref{eq:second} is by considering its linearized version, as follows:
\begin{equation}\label{eq:second:lin}
   c \, \Delta \ddot {y} + d \, \Delta\dot {y} + k {\Delta y} = 0 \, .
\end{equation}
The variations of stored energy ($\Delta E$) and power dissipation ($\Delta P_l$) of the oscillator are then \cite{chen2012energy}:
\begin{equation}\label{eq:KEDP}
 \Delta E = \frac{1}{2} \, c \, \Delta \dot {y}^2 \ , 
  \hspace{3mm}
  \Delta P_l =  d \, \Delta \dot {y}^2 \, , 
\end{equation}
while its eigenvalues are $\lambda = ({-d \pm \sqrt{d^2-4ck}})/{2c} $
and, thus, the following relationship holds:
\begin{equation}\label{eq:link}
\frac{\Delta E}{\Delta P_l} 
= \frac{c}{2d} =
- \frac{1}{4 \, \Re\{\lambda\}} \, .
\end{equation}
From \eqref{eq:link}, it is clear that dynamics faster than the time scale of interest are likely to lead to high damping and also violate the requirement for available stored energy, as fast eigenvalues are in general linked to lower amounts of energy and higher damping ratios.

\subsection{Response to Short-Circuits}
\label{sec:ssc}

The short-circuit current that a \ac{sm} can tolerate before protections are activated is multiple of the nominal for some time.  The high ``thermal inertia" of \acp{sm} is in contrast to the limited ability to overload power converters. This implies that, even if a \ac{cig} is controlled to reproduce well the response of a \ac{sm} under small disturbances, the same can not be achieved during severe voltage drops, unless the converter design is significantly overrated (e.g.~by 6 to 7 times).  However, such a design is not practical for economical reasons.  This appears to be a rather severe limitation of \acp{vsm} in general, given that replication of the behavior of \acp{sm} is of upmost importance during large disturbances such as faults.

\subsection{Remarks}

The following remarks are relevant:
\begin{itemize}
\item The conditions discussed above focus mainly on the critical (for low-inertia systems) time scale of the \ac{sm} inertial response, which is also the most relevant time scale for the emulation of \ac{sm} dynamics.  Slower actions, including primary and secondary frequency regulation, are not a concern, since they can be conveniently implemented with standard controllers without the need to make any analogy with a \ac{sm}.
\item In the classical model \eqref{eq:motion} the internal emf $e'$ is constant, which makes the \ac{sm} an ideal voltage source.  In normal operating conditions, \acp{sm} are not ideal sources but do regulate the voltage magnitude at their terminal bus.  This capability is not intrinsic of the \ac{sm} \textit{per se}, e.g., permanent-magnet \acp{sm} are unable to provide this control.  On the other hand, in practice, \ac{sm}-emulating devices are expected to provide voltage regulation.  This is the case of \acp{cig} controlled through a grid-forming strategy, or more precisely, voltage-forming current-following control \cite{li2022revisiting}.  We note, however, that the specific property of the \acp{sm} is to be \textit{frequency-forming} -- which is a consequence of \eqref{eq:motion} -- not necessarily voltage-forming.
\end{itemize}

\section{Case Study}
\label{sec:case}

In this section we discuss through numerical simulations the behavior of devices that have been proposed in the literature as analogous and/or equivalent to \acp{sm}.  Section~\ref{sec:example}
is based on the simplified model \eqref{eq:second}, while Section~\ref{sec:9bus} is based on the well-known WSCC 9-bus test system.

\subsection{Illustrative Example}
\label{sec:example}

In this section we consider different devices modeled as second-order oscillators in the form of \eqref{eq:second}.
The first device is a conventional \ac{sm}.  The second device is a droop-based control that acts in the time scale of the \ac{sm} inertial response.  Since droop controls are in general not oscillatory, modeling such device with \eqref{eq:second} implies that the oscillator is overdamped, or equivalently, that $d$ is relatively large.  We note that energy availability is not a given for droop controllers.  Assuming a droop control combined with sufficient power reserve availability that can be used very fast following a disturbance is under certain conditions what has been often defined in the recent literature as \ac{vsm}.  
The last device considered is a simplified \ac{pll}. The \ac{pll} is much faster than a \ac{sm} and also does not have the required energy to provide inertial response. On the other hand, a \ac{pll} may oscillate, although the damping ratio of \ac{pll} oscillations is not necessarily similar to that of \ac{sm} oscillations.
Table~\ref{tab:conditions} summarizes how droop control, \ac{vsm}, and \ac{pll} compare to a conventional \ac{sm} connected to a power system in view of the conditions for energy availability, time scale, damping, and short-circuit response, discussed in Section~\ref{sec:conditions}. 

\begin{table}[ht!]
  \centering
  \caption{Comparison of droop control, \ac{vsm}, and \ac{pll},
  with conventional \ac{sm}.}
    \renewcommand{\arraystretch}{1.1}
    \setlength{\tabcolsep}{4pt}
  \label{tab:conditions}
  \centering 
  \begin{minipage}{0.5\textwidth}
  \centering 
  \begin{threeparttable}
    \begin{tabular}{lcccc}
      \toprule
     {Device} & {Energy} & 
      {{Time scale}} & {{Damping}} 
      & Short-circuit response 
      \\
      \midrule
        Conventional \ac{sm} & \cmark & \cmark & \cmark & \cmark 
      \\
       {Droop control} & \xmark & \cmark & \xmark  & \xmark 
      \\
      \ac{vsm} & \cmark
      & \cmark &  \xmark
      & \xmark  
      \\
      \ac{pll} & \xmark  & \xmark & \xmark & \xmark 
      \\
      \bottomrule
    \end{tabular}
  \end{threeparttable}
  \end{minipage}
\end{table}

\begin{figure}[ht!]
  \centering
  \begin{subfigure}{.9\linewidth}
    \centering
    \resizebox{\linewidth}{!}{\includegraphics{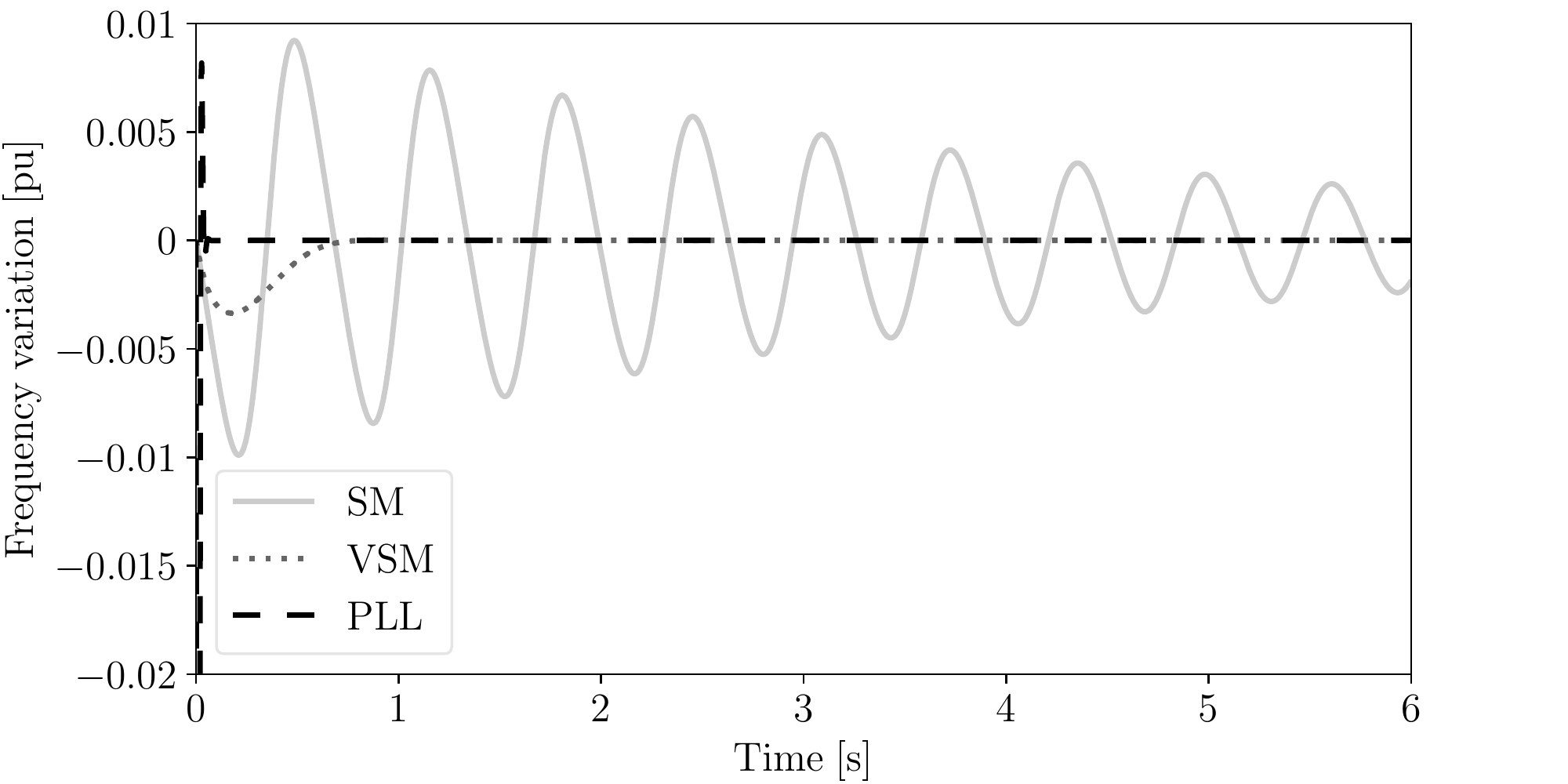}}
    \label{fig:w}
     \vspace{-3.5mm}
  \end{subfigure}    
  \vspace{-4.5mm}
  \begin{subfigure}{.9\linewidth}
    \centering
    \resizebox{\linewidth}{!}{\includegraphics{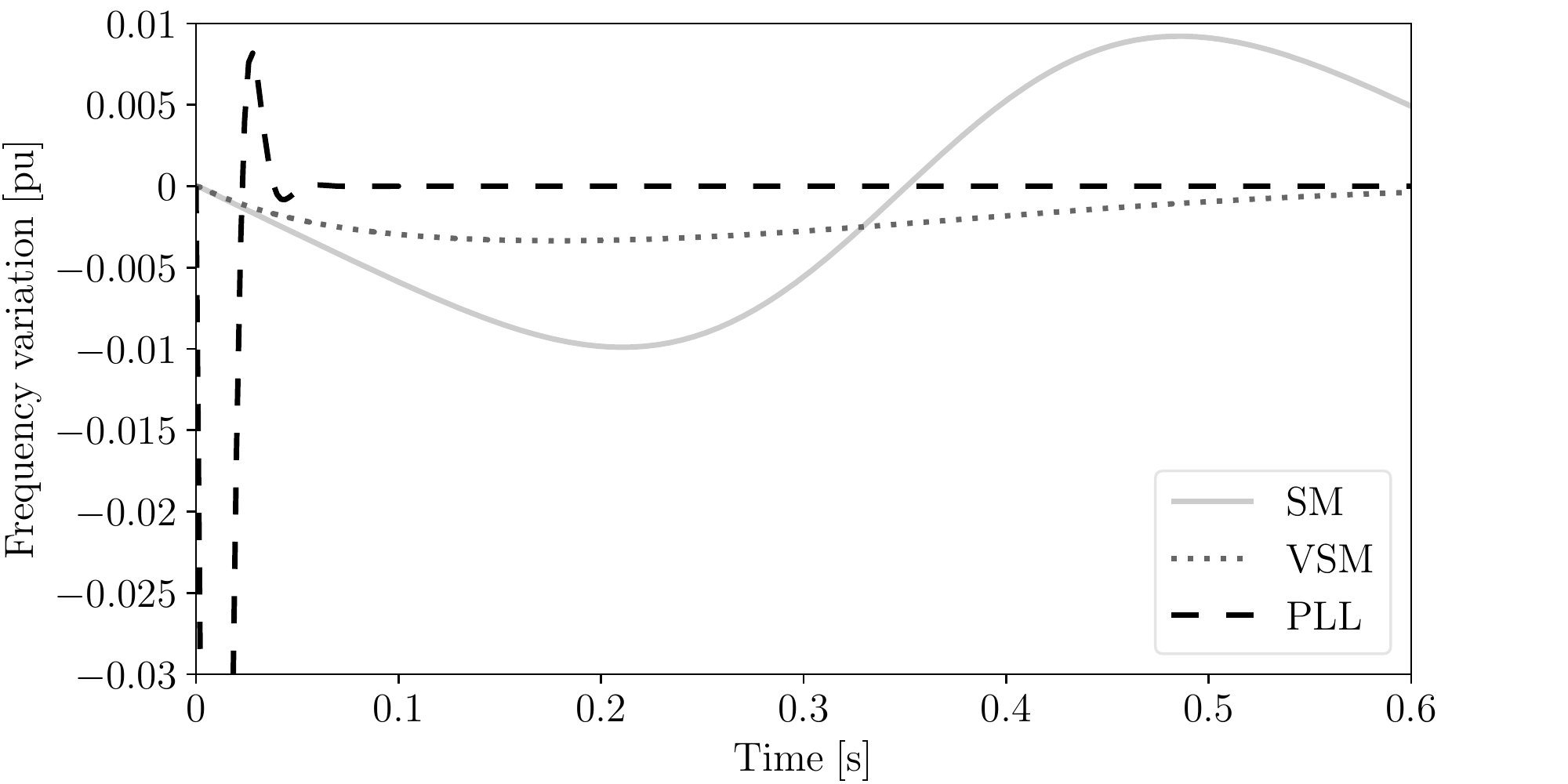}}
    \label{fig:w_z}
  \end{subfigure}
  \caption{Response of \ac{sm}, \ac{vsm}, and \ac{pll}.}
  \label{fig:tds}
\end{figure}

Figure~\ref{fig:tds} shows how the step responses of the \ac{sm}, \ac{vsm}, and \ac{pll}, compare to each other.  The top panel shows the frequency variations of the devices, while the bottom provides a close-up of the same plot.  The values of $c$ and $d$ used for each device are given in Table~\ref{tab:params}.  These values yield the following ratios between stored energy and power dissipation in \eqref{eq:link} for the three devices:
\begin{itemize}
\item \ac{sm}: \hspace{2cm} ${\Delta E}/{\Delta P_l} = 1$ .
\item \ac{vsm}: \hspace{1.75cm} ${\Delta E}/{\Delta P_l} = 0.03$ .
\item \ac{pll}: \hspace{1.85cm} ${\Delta E}/{\Delta P_l} = 0.00167$ .
\end{itemize}
The ${\Delta E}/{\Delta P_l} $ ratio for the \ac{sm} is two and three, respectively, orders of magnitude larger than those of the \ac{vsm} and \ac{pll}.  This result, as well as the plots in Fig.~\ref{fig:tds}, are illustrative of the significant difference between the conventional behavior of \acp{sm} in power systems and the responses of devices that do not satisfy the constraints described in Section~\ref{sec:conditions}.  
\begin{table}[ht!]
  \centering
  \caption{Parameters of second-order oscillators.}
  \renewcommand{\arraystretch}{1.1}
  \label{tab:params}\centering 
  \begin{threeparttable}
    \begin{tabular}{l|ccc}
      \toprule
      Device & \ac{sm} & \ac{vsm} & \ac{pll} \\
      \midrule
      $c$ & $6$ & $6$ & $0.01$ \\
      $d$ & $3$ & $100$ & $3$ \\ 
      \bottomrule
    \end{tabular}
  \end{threeparttable}
\end{table}

\subsection{WSCC 9-Bus System}
\label{sec:9bus}

\begin{figure}[ht!]
  \centering
  \begin{subfigure}{.99\linewidth}
    \centering
    \resizebox{\linewidth}{!}{\includegraphics{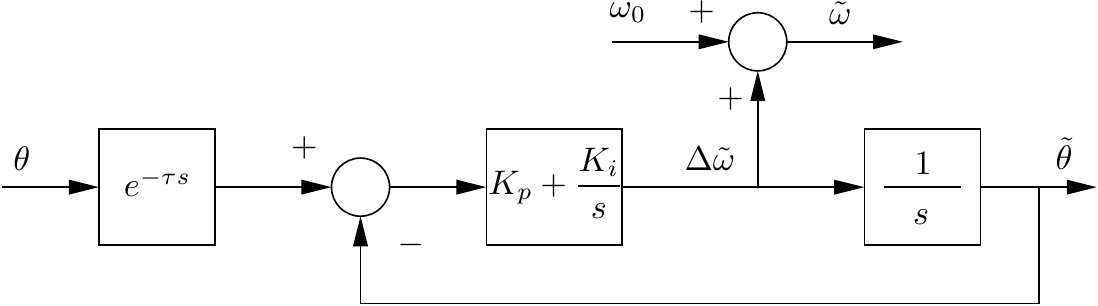}}
    \vspace{-3.2mm}
    \caption{Synchronous reference frame \ac{pll}.}
    \label{fig:wscc:tds:pll}
     \vspace{3.2mm}
  \end{subfigure}    
  \begin{subfigure}{.9\linewidth}
    \centering
    \resizebox{\linewidth}{!}{\includegraphics{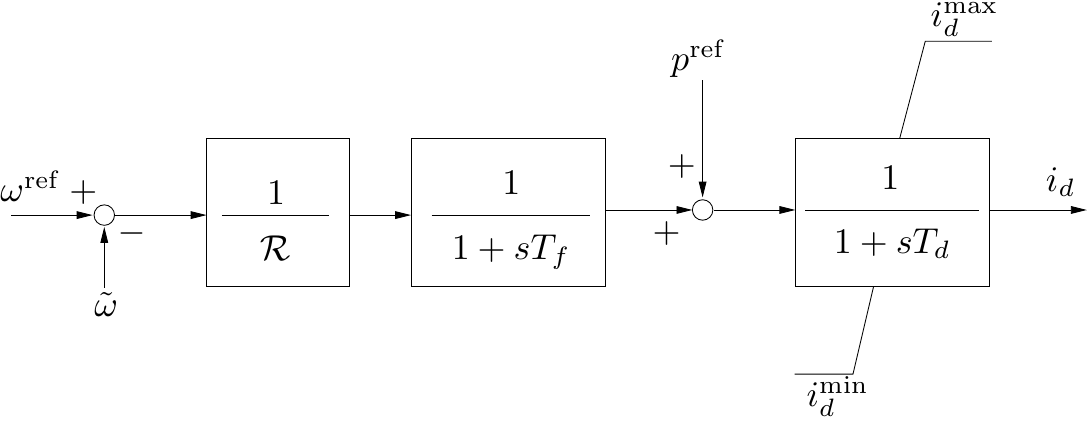}}
    \vspace{-3.2mm}
    \caption{Frequency control.}
    \label{fig:wscc:tds:fc}
  \end{subfigure}
  \caption{\ac{cig} synchronization and frequency control loops.}
  \label{fig:wscc:cig}
\end{figure}

\begin{figure}[ht!]
  \centering
  \resizebox{0.94\linewidth}{!}{\includegraphics{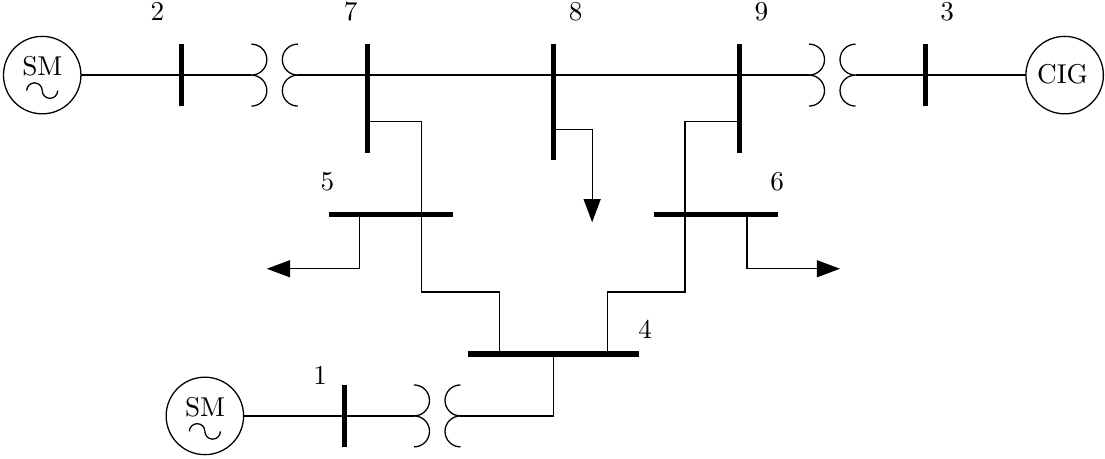}}
  \caption{Modified WSCC $9$-bus test system.}
  \label{fig:wscc}
  \vspace{-0.75mm}
\end{figure}

This section is based on the WSCC 9-bus test system. The network comprises six transmission lines and three medium voltage/high voltage transformers; during transients, loads are modeled as constant admittances; two \acp{sm} are connected to buses~1 and 2, while, for the needs of this paper, the \ac{sm} at bus~3 is replaced by a \ac{cig}.  The modified test system is shown in Fig.~\ref{fig:wscc}.  The \ac{cig} at bus~3 synchronizes to the power grid through a synchronous reference frame \ac{pll} and provides primary frequency response through a droop-based controller that receives the error between the reference and estimated by the \ac{pll} frequency $\omega^{\rm ref}-\tilde \omega$ and regulates the $d$-axis current component $i_d$ in the $dq$-reference frame \cite{zhong2022improving}.  The estimated by the \ac{pll} frequency is obtained through a proportional-integral control whose input is the error between the measured and the estimated phase angles $\theta-\tilde\theta$.  The block diagrams and parameter values of the \ac{pll} and frequency control models are presented in Fig.~\ref{fig:wscc:cig} and Table~\ref{tab:cigpars}.

\begin{table}[ht!]
  \centering
  \renewcommand{\arraystretch}{1.1}
  \caption{\ac{cig} control parameters.}
  \label{tab:cigpars}
  \begin{threeparttable}
    \begin{tabular}{p{1.5cm}l} 
      \toprule
      \ac{pll} & $K_p=0.1$, $K_i=0.05$\\
      Frequency  & $\mathcal{R} = 0.05$, $T_{f} = 1.2$~s,  $T_{d}=0.6$~s \\
      \bottomrule
    \end{tabular}
  \end{threeparttable}
  \vspace{-3mm}
\end{table}

The dynamic order of the \ac{cig} model, including the droop-based control, $d$ and $q$ axis current control, and \ac{pll} dynamics, is 5.  The dynamic order of each \ac{sm}, including frequency and voltage regulators, is 11.

Small-signal stability analysis shows that the system's electromechanical oscillation is represented by the complex pair of eigenvalues $-0.2135 \pm \jj 8.6897$, with natural frequency $1.38$~Hz and damping ratio $2.46$\%.  Comparison of the modal participation, see \cite{book:eigenvalue}, in this mode of the \ac{cig} droop-based control state variable, the \ac{cig}'s \ac{pll} angle estimation state variable, as well as of the rotor speeds of the \acp{sm} at buses~1~(SM1) and 2~(SM2), is provided in Table~\ref{tab:pf}.  These results indicate that, as expected, the \ac{pll} and droop control variables are decoupled from the electromechanical mode.
\begin{table}[ht!]
  \centering
  \caption{Participation factors for electromechanical mode.}
  \renewcommand{\arraystretch}{1.1}
  \label{tab:pf}\centering 
  \begin{threeparttable}
    \begin{tabular}{lccccc}
      \toprule
      Variable & Participation factor \\
      \midrule
      Rotor speed of SM1 & $0.16$ \\
      Rotor speed of SM2 & $0.32$ \\
      \ac{cig} droop control state & $0.00$ \\
      \ac{pll} angle estimation state & $0.00$ \\
      \bottomrule
    \end{tabular}
  \end{threeparttable}
\end{table}

\begin{figure}[ht!]
  \centering
  \begin{subfigure}{.9\linewidth}
    \centering
    \resizebox{\linewidth}{!}{\includegraphics{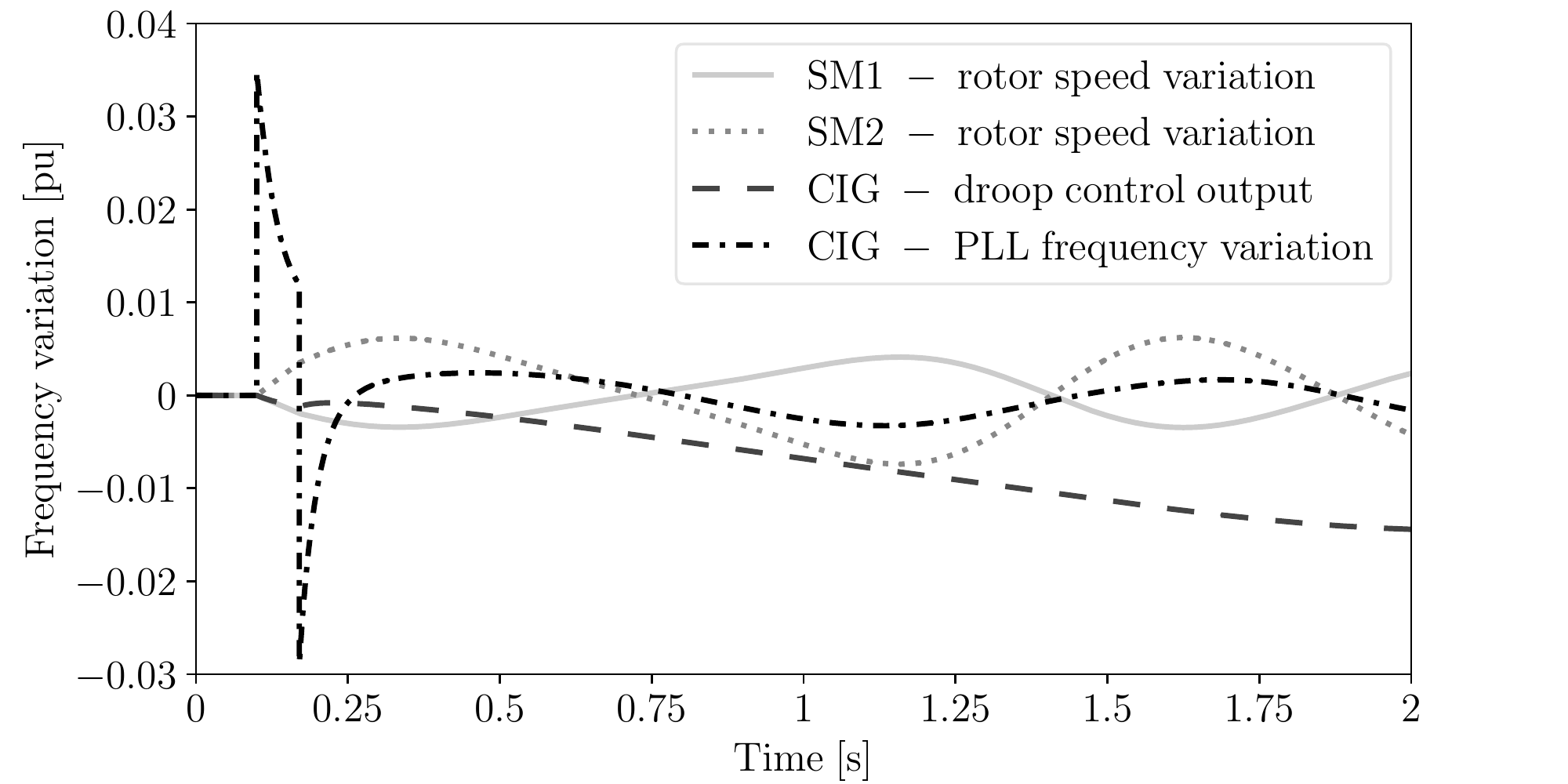}}
    \vspace{-3.2mm}
    \caption{Frequency variation.}
    \label{fig:wscc:tds:w}
     \vspace{3.2mm}
  \end{subfigure}    
  \begin{subfigure}{.9\linewidth}
    \centering
    \resizebox{\linewidth}{!}{\includegraphics{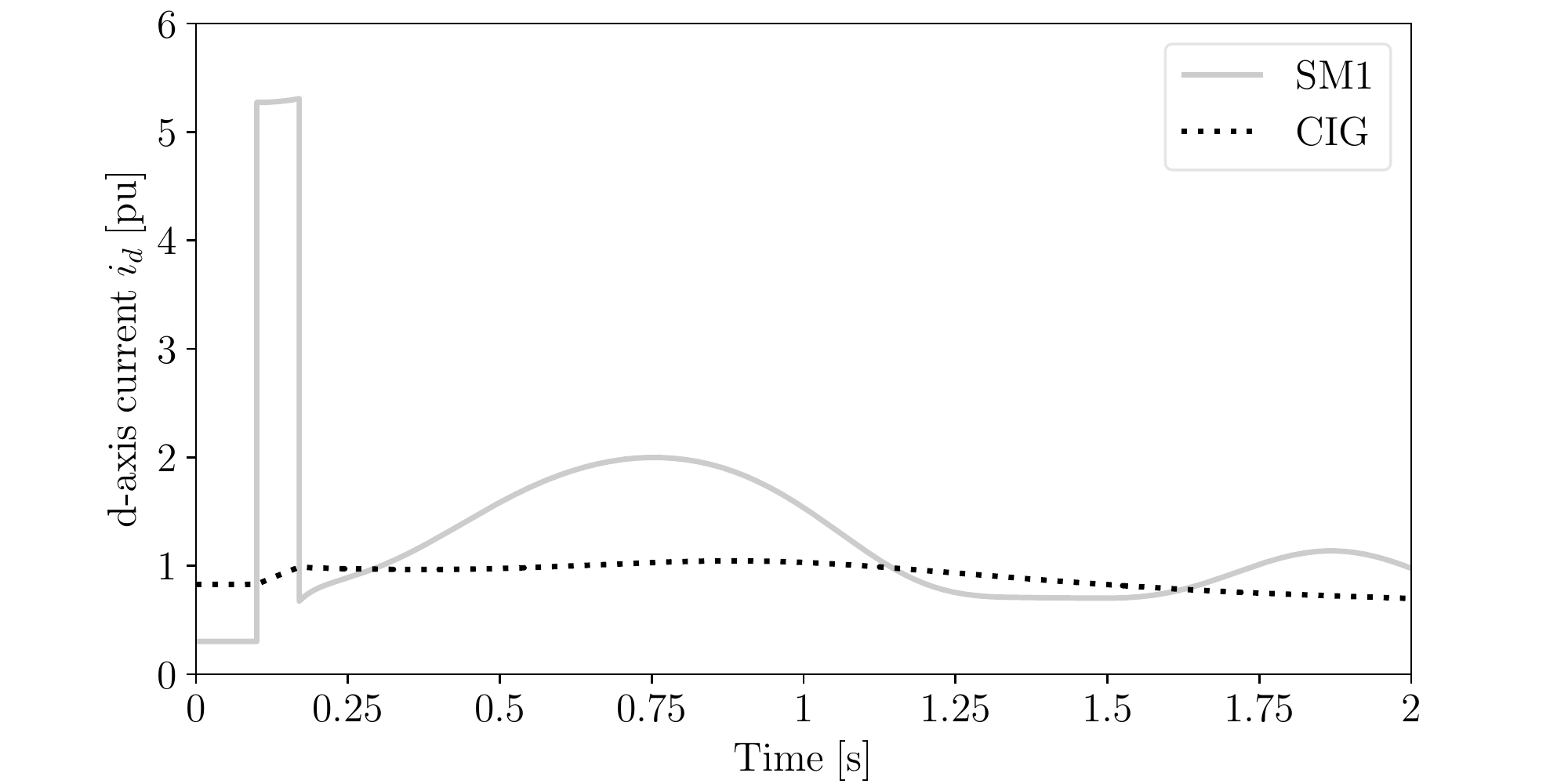}}
    \vspace{-3.2mm}
    \caption{d-axis current.}
    \label{fig:wscc:tds:id}
  \end{subfigure}
  \caption{Response following the three-phase fault, 9-bus system.}
  \label{fig:wscc:tds}
\end{figure}

We consider a three-phase fault at bus~5 at $t=0.1$~s. The fault is cleared after $70$~ms, by tripping the line that connects buses~5 and 7.  Results are summarized in Fig.~\ref{fig:wscc:tds}.  In particular, Fig.~\ref{fig:wscc:tds:w} shows the speed variation of the \acp{sm}; the (normalized) output of the \ac{cig} droop control; and the frequency variation at bus~3 as estimated by the \ac{pll}.  Figure~\ref{fig:wscc:tds:id} illustrates the tight limit in the ability to overload the \ac{cig} and thus to support the system during the fault, by comparing the d-axis current component of the stator of the \ac{sm} at bus~1 to the d-axis regulated current component of the \ac{cig}.  Once again, results are representative of the large qualitative deviations between the behavior of a conventional \ac{sm} connected to a power system and devices that are not designed to resemble its dynamics according to the conditions discussed in Section~\ref{sec:conditions}.

\section{Conclusions}
\label{sec:conclusion}

The paper shows that a given second-order oscillatory device resembles the dynamic response of a \ac{sm} only if it satisfies certain conditions.  These conditions are concerned with the device's availability of energy, time scale of action, damping of oscillations, and response during short-circuits.  Devices that do not fulfill these conditions have been characterized in the recent literature as equivalent or analogous to \acp{sm}.  Such devices should not be confused with and/or misinterpreted as replicating the traditional behavior of a \ac{sm} connected to a power network.



\end{document}